\documentclass[twocolumn]{jpsj2}

\def\runauthor{%
J. \textsc{Otsuki}, H. \textsc{Kusunose} and Y. \textsc{Kuramoto}
}

\title{%
Self-Consistent Perturbation Theory for 
Thermodynamics of\\ Magnetic Impurity Systems
}

\author{%
Junya \textsc{Otsuki}\thanks{E-mail address: otsuki@cmpt.phys.tohoku.ac.jp}
,Hiroaki \textsc{Kusunose} and Yoshio \textsc{Kuramoto}
}

\inst{%
Department of Physics, Tohoku University, Sendai 980-8578
}

\recdate{\today}

\abst{%
Integral equations for thermodynamic quantities are derived in the framework of the non-crossing approximation (NCA). 
Entropy and specific heat of $4f$ contribution are calculated without numerical differentiations of thermodynamic potential. 
The formulation is applied to systems such as PrFe$_4$P$_{12}$ with singlet-triplet crystalline electric field  (CEF) levels.
}

\kword{%
thermodynamics, entropy, specific heat, non-crossing approximation (NCA), crystalline electric field (CEF), filled skutterudite, PrFe$_4$P$_{12}$
}

\begin{document}
\maketitle

The resolvent method has been developed as a powerful technique for the Anderson model
with strong correlations \cite{nca1, grewe, keiter-morandi, bickers}. It enable us to derive dynamics as well as thermodynamics. 
The lowest-order self-consistent approximation is called the NCA, 
where the self-energy part of the resolvent is determined by 
the lowest order skeleton diagram.
Let us first consider the simplest case where $4f^0$ and $4f^1$ configurations are relevant.  
Once the resolvent $R_\gamma (z)$ for each $4f$ state $\gamma$
is obtained, the partition function 
$Z_f$ of $4f$ part is given by
\cite{nca1, grewe, keiter-morandi, bickers}
\begin{align}
	Z_f = \int \text{d}\epsilon\ \text{e}^{-\beta \epsilon} \sum_{\gamma} \eta_{\gamma}(\epsilon),
	\label{eq:part_func_integ}
\end{align}
where 
$\eta_{\gamma}(\epsilon) = -\text{Im} R_{\gamma}(\epsilon +\text{i}\delta)/\pi$ with $\delta$ being positive infinitesimal is a spectral function. 
Physical quantities such as single-particle spectra and the thermodynamic potential $\Omega_f = -T\ln Z_f$ of $4f$ contribution are derived with use of resolvents. 

On the other hand, entropy $S_f = -\partial \Omega_f / \partial T$ and specific heat $C_f = T \partial S_f / \partial T$ have been computed by numerical 
differentiation of $\Omega_f$ with respect to temperature $T$. 
The differentiation requires one to calculate $Z_f$ at numerous values of temperature.  To obtain reasonable numerical accuracy, substantial time is required in computations, especially for the second derivatives.
Hence a method is desirable
that yields the $T$-derivatives of $\Omega_f$ at a given value of temperature without numerical differentiation. 
In this paper, we present a method 
to compute the $T$-derivatives of spectral functions by solving integral equations, instead of taking the numerical derivatives.

In order to establish notations for our development, we briefly summarize the NCA integral equations and their equivalents.
The NCA integral equations for the self-energies 
$\Sigma_{\gamma} (z)$
of resolvents are given by \cite{nca1, grewe, keiter-morandi, bickers}
\begin{align}
	\Sigma_{0} (z) &= \int \text{d}\epsilon W(\epsilon) f(\epsilon) \sum_{\beta} R_{\beta}(z+\epsilon), 
\label{eq:self0}
	\\
	\Sigma_{1} (z) &= \int \text{d}\epsilon W(\epsilon) [1-f(\epsilon)] R_0 (z-\epsilon),
\label{eq:self}
\end{align}
where $\beta$ denotes one of $4f^1$ states, 
$f(\epsilon)$ is the Fermi distribution function, and 
$W(\epsilon)$ is the product of hybridization squared and density of states of conduction electron. 
Since the Boltzmann factor becomes singular at low temperatures, an auxiliary spectral function 
$\xi_{\gamma}(\epsilon) = Z_f^{-1} \text{e}^{-\beta \epsilon} \eta_{\gamma}(\epsilon)$ has been introduced \cite{nca3}. 
The quantity stands for intensity of removal of state $\gamma$.
Alternatively, $\xi_{\gamma}(\epsilon)$ is interpreted as a result of the operation
$\mathcal{P}$, which is defined by 
$\mathcal{P}R(\omega)=Z_f^{-1} \text{e}^{-\beta \epsilon} (-)\pi^{-1}\text{Im}R(\omega+\text{i}\delta)$.
Similar operation gives 
$\sigma_{\gamma}(\omega)=\mathcal{P}\Sigma_{\gamma}(\omega)$.
Then we obtain equations equivalent to the NCA equations: 
\begin{align}
	\xi_{\gamma} (\omega) &= |R_{\gamma}(\omega+\text{i}\delta)|^2 \sigma_\gamma(\omega),
	\label{eq:xi} \\
	\sigma_{0} (\omega) &= \int \text{d}\epsilon W(\epsilon) [1-f(\epsilon)] \sum_{\beta} \xi_{\beta}(\omega+\epsilon), \\
	\sigma_{1} (\omega) &= \int \text{d}\epsilon W(\epsilon) f(\epsilon) \xi_0 (\omega-\epsilon),
	\label{eq:sigma1}
\end{align}
which gives $\xi_{\gamma}(\epsilon)$ without manipulation of the Boltzmann factor.
Although 
linear equations (\ref{eq:xi})--(\ref{eq:sigma1}) do not determine the norms of $\xi_{\gamma}(\omega)$ and $\sigma_{\gamma}(\omega)$,  they are determined by the following sum-rule:
\begin{align}
	\int \text{d}\omega \sum_{\gamma} \xi_{\gamma} (\omega) = 1,
	\label{eq:sum_rule}
\end{align}
which are obtained with use of eq. (\ref{eq:part_func_integ}). 


Now we derive new integral equations for thermodynamics.
For any energy $\epsilon$, 
$Z_f$ is represented in terms of spectral intensities as
\begin{align}
	Z_f = \text{e}^{-\beta \epsilon} \eta_{\gamma}(\epsilon) / \xi_{\gamma}(\epsilon).
	\label{eq:part_func}
\end{align}
Since $\eta_{\gamma}(\epsilon)$ and $\xi_{\gamma}(\epsilon)$ are obtained by different integral equations, 
eq. (\ref{eq:part_func}) yields $Z_f$.
Equation (\ref{eq:part_func}) leads to an expression of entropy in terms of spectral functions of a resolvent
\begin{align}
	S_f
	 = \ln \frac{\eta_{\gamma}(\epsilon)}{\xi_{\gamma}(\epsilon)}
	 + T \left[ \frac{1}{\eta_{\gamma}(\epsilon)} \frac{\partial \eta_{\gamma}(\epsilon)}{\partial T}
	 - \frac{1}{\xi_{\gamma}(\epsilon)} \frac{\partial \xi_{\gamma}(\epsilon)}{\partial T} \right],
\end{align}
with arbitrary $\gamma$. Thus $T$-derivatives of $\eta_{\gamma}(\epsilon)$ and $\xi_{\gamma}(\epsilon)$ are required.

%
Performing derivatives with respect to $T$ in eqs. (\ref{eq:self0}) and  (\ref{eq:self}), 
we obtain a set of integral equations for the first derivatives:
\begin{align}
	\frac{\partial R_{\gamma}(z)}{\partial T}
	 &= R_{\gamma}(z)^2 \frac{\partial \Sigma_{\gamma}(z)}{\partial T}, 
\label{eq:resolv_deriv1} \\
	\frac{\partial \Sigma_{0} (z)}{\partial T}
	 &= \int \text{d}\epsilon W(\epsilon) f(\epsilon) \sum_{\beta} \frac{\partial R_{\beta}(z+\epsilon)}{\partial T} + F_0 (z), \\
	\frac{\partial \Sigma_{1} (z)}{\partial T}
	 &= \int \text{d}\epsilon W(\epsilon) [1-f(\epsilon)] \frac{\partial R_0 (z-\epsilon)}{\partial T} + F_1(z),
\label{eq:self_deriv1}
\end{align}
where functions $F_0(z)$ and $F_1(z)$ are defined as
\begin{align}
	F_0(z) &= \int \text{d}\epsilon W(\epsilon) \frac{\partial f(\epsilon)}{\partial T} \sum_{\beta} R_{\beta}(z+\epsilon), \\
	F_1(z) &= -\int \text{d}\epsilon W(\epsilon) \frac{\partial f(\epsilon)}{\partial T} R_0 (z-\epsilon),
\end{align}
which can be computed from resolvents.
Numerical iterations of eqs. (\ref{eq:resolv_deriv1})--(\ref{eq:self_deriv1}) yield $T$-derivative of resolvents.


We now proceed to derivation of $\partial \xi_{\gamma}(\epsilon) / \partial T$ from integral equations.
Performing $T$-derivatives in eqs. (\ref{eq:xi})--(\ref{eq:sigma1}), we obtain equations for $\partial \xi_{\gamma}(\epsilon) / \partial T$ as
\begin{align}
	\frac{\partial \xi_{\gamma} (\omega)}{\partial T} &= |R_{\gamma}(\omega +\text{i}\delta)|^2 \frac{\partial \sigma_\gamma(\omega)}{\partial T} + g_\gamma(\omega),
	\label{eq:xi_dT} \\
	\frac{\partial \sigma_{0} (\omega)}{\partial T}
	 &= \int \text{d}\epsilon W(\epsilon) [1-f(\epsilon)] \sum_{\beta} \frac{\partial \xi_{\beta}(\omega+\epsilon)}{\partial T} + G_0 (\omega), \\
	\frac{\partial \sigma_{1} (\omega)}{\partial T}
	 &= \int \text{d}\epsilon W(\epsilon) f(\epsilon) \frac{\partial \xi_0 (\omega-\epsilon)}{\partial T} + G_1(\omega),
	 \label{eq:sigma1_dT}
\end{align}
where auxiliary functions are defined by
\begin{align}
	g_\gamma(\omega) &= 2\text{Re} \left[ \frac{\partial R_{\gamma}(\omega +\text{i}\delta)}{\partial T} R_{\gamma} (\omega +\text{i}\delta)^* \right] \sigma_\gamma(\omega), \\
	G_0 (\omega) &= -\int \text{d}\epsilon W(\epsilon) \frac{\partial f(\epsilon)}{\partial T} \sum_{\beta} \xi_{\beta}(\omega+\epsilon), \\
	G_1 (\omega) &= \int \text{d}\epsilon W(\epsilon) \frac{\partial f(\epsilon)}{\partial T} \xi_0 (\omega-\epsilon).
\end{align}
Since eqs. (\ref{eq:xi_dT})--(\ref{eq:sigma1_dT}) coincide with eqs. (\ref{eq:xi})--(\ref{eq:sigma1}) in the case of vanishing functions $G_{\gamma}(\omega)$ and $g_{\gamma}(\omega)$, 
general solutions are given in terms of  an arbitrary parameter $c$ as
 $\partial \xi_{\gamma}(\omega)/\partial T + c \xi_{\gamma}(\omega)$ and $\partial \sigma_{\gamma}(\omega)/\partial T + c \sigma_{\gamma}(\omega)$.
The physical value for $c$ is fixed by 
the following condition:
\begin{align}
	\int \text{d}\omega \sum_{\gamma} \frac{\partial \xi_{\gamma} (\omega)}{\partial T} = 0,
	\label{eq:sum_rule_dT1}
\end{align}
which follows 
from the sum rule given by eq. (\ref{eq:sum_rule}).

Equations for the second derivatives are derived in a manner similar to the first derivatives.
Specific heat $C_f$ of $4f$ contribution is represented in terms of spectral functions as 
\begin{align}
	C_f 
	&= 2T \left[ \frac{1}{\eta_{\gamma}(\epsilon)} \frac{\partial \eta_{\gamma}(\epsilon)}{\partial T}
	 - \frac{1}{\xi_{\gamma}(\epsilon)} \frac{\partial \xi_{\gamma}(\epsilon)}{\partial T} \right] \nonumber \\
	 &-T^2 \left[ \frac{1}{\eta_{\gamma}(\epsilon)^2} \left( \frac{\partial \eta_{\gamma}(\epsilon)}{\partial T} \right)^2
	 - \frac{1}{\xi_{\gamma}(\epsilon)^2} \left( \frac{\partial \xi_{\gamma}(\epsilon)}{\partial T} \right)^2 \right] \nonumber \\
	 &+T^2 \left[ \frac{1}{\eta_{\gamma}(\epsilon)} \frac{\partial^2 \eta_{\gamma}(\epsilon)}{\partial T^2}
	 - \frac{1}{\xi_{\gamma}(\epsilon)} \frac{\partial^2 \xi_{\gamma}(\epsilon)}{\partial T^2} \right].
\end{align}
We note that $C_f$ here is the specific heat with the chemical potential fixed.  
A set of integral equations for the second derivatives of resolvents are obtained by following modifications to eqs. (\ref{eq:resolv_deriv1})--(\ref{eq:self_deriv1}): addition of a term $\tilde{f}_{\gamma}(z)$ to the right hand side of eq. (\ref{eq:resolv_deriv1}), and replacement of $\partial R_{\gamma} /\partial T$, $\partial \Sigma_\gamma /\partial T$ and $F_\gamma$ by $\partial^2 R_{\gamma} /\partial T^2$, $\partial^2 \Sigma_\gamma /\partial T^2$ and $\tilde{F}_\gamma$, respectively. 
$\tilde{F}_\gamma(z)$ and $\tilde{f}_{\gamma}(z)$ are computed with use of the zeroth and first derivatives of resolvent as follows:
\begin{align}
	\tilde{F}_0 (z) &= \int \text{d}\epsilon W(\epsilon) \sum_{\beta} \nonumber \\
	 \times& \left[
	 2\frac{\partial f(\epsilon)}{\partial T} \frac{\partial R_{\beta} (z+\epsilon)}{\partial T}
	 +\frac{\partial^2 f(\epsilon)}{\partial T^2} R_{\beta} (z+\epsilon) \right], \\
	\tilde{F}_1 (z) &= -\int \text{d}\epsilon W(\epsilon) \nonumber \\
	 \times& \left[
	 2\frac{\partial f(\epsilon)}{\partial T} \frac{\partial R_0 (z-\epsilon)}{\partial T}
	 +\frac{\partial^2 f(\epsilon)}{\partial T^2} R_0 (z-\epsilon) \right], \\
	\tilde{f}_{\gamma}(z) &= 2R_{\gamma}(z) \frac{\partial R_{\gamma}(z)}{\partial T}
	 \frac{\partial \Sigma_{\gamma}(z)}{\partial T}.
\end{align}
Equations for $\partial^2 \xi/\partial T^2$ are obtained by replacement of $\partial \xi_{\gamma} /\partial T$, $\partial \sigma_{\gamma} /\partial T$, $G_\gamma$ and $g_{\gamma}$ by $\partial^2 \xi_{\gamma} /\partial T^2$, $\partial^2 \sigma_{\gamma} /\partial T^2$, $\tilde{G}_\gamma$ and $\tilde{g}_{\gamma}$ in eqs. (\ref{eq:xi_dT})--(\ref{eq:sigma1_dT}), respectively. 
Auxiliary functions $\tilde{G}_\gamma(\omega)$ and $\tilde{g}_{\gamma}(\omega)$ are defined by
\begin{align}
	\tilde{G}_0 (\omega) &= -\int \text{d}\epsilon W(\epsilon) \sum_{\beta} \nonumber \\
	 \times& \left[
	 2\frac{\partial f(\epsilon)}{\partial T} \frac{\partial \xi_{\beta} (\omega+\epsilon)}{\partial T}
	 +\frac{\partial^2 f(\epsilon)}{\partial T^2} \xi_{\beta} (\omega+\epsilon) \right], \\
	\tilde{G}_1 (\omega) &= \int \text{d}\epsilon W(\epsilon) \nonumber \\
	 \times& \left[
	 2\frac{\partial f(\epsilon)}{\partial T} \frac{\partial \xi_0 (\omega-\epsilon)}{\partial T}
	 +\frac{\partial^2 f(\epsilon)}{\partial T^2} \xi_0 (\omega-\epsilon) \right], \\
	\tilde{g}_{\gamma}(\omega) &= 2\frac{\partial |R_{\gamma}(\omega+\text{i}\delta)|^2}{\partial T} \frac{\partial \sigma_{\gamma}(\omega)}{\partial T} \nonumber \\
	 &+ \frac{\partial^2 |R_{\gamma}(\omega+\text{i}\delta)|^2}{\partial T^2} \sigma_{\gamma}(\omega).
\end{align}
The particular solution with a proper coefficient corresponding to $c$ used for the first derivatives can be extracted in a similar way.

We apply the above formation to the CEF singlet-triplet system taking PrFe$_4$P$_{12}$ as a target. 
Considering hybridization between $4f^2$ and conduction electrons of the molecular orbital with $a_u$ symmetry, which only has spin degrees of freedom, exchange interaction is written as\cite{otsuki1}
\begin{align}
	H_{\text{s-t}} &= 
	\Delta_{\rm CEF} \mib{S}_1\cdot \mib{S}_2 +
	(J_1\mib{S}_1 + J_2\mib{S}_2)\cdot \mib{s}_c,
\label{eq:st-kondo}
\end{align}
where the CEF singlet-triplet states are represented by dimer of two pseudo-spins $\mib{S}_1$ and $\mib{S}_2$ \cite{shiina-aoki}, and $\Delta_{\rm CEF}$ is the CEF splitting.
We have derived integral equations for resolvent and dynamical quantities in the NCA\cite{otsuki2}. Its application to thermodynamics in the above framework is more complicated but straightforward. We shall show only the numerical results for entropy and specific heat. 
It is known that the NCA is justified in the large $n$ limit, where $n$ is scattering channel. 
In this model, however, conduction electron only has spin degeneracy, i.e., $n=2$, whereas local $4f^2$ configuration has four levels. 
Hence derivation of the accurate energy scale including the Kondo effect cannot be expected in the NCA.
If one can derive the accurate scale by another method such as the numerical renormalization group (NRG), however, one may hope that the temperature dependence can be reasonably understood by rescaling the NCA results.
%

\begin{figure}[t]
	\begin{center}
	\includegraphics[width=0.95\linewidth]{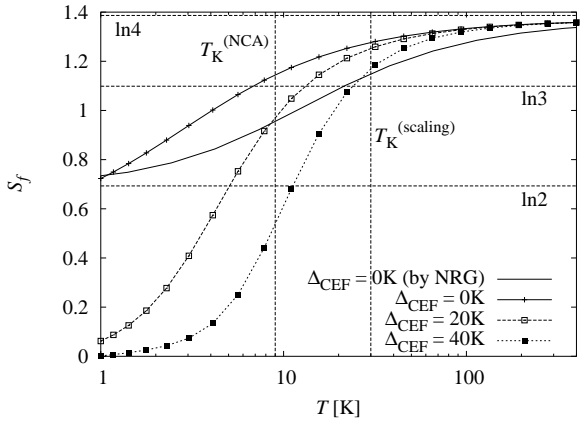}
	\end{center}
	\caption{Entropy $S_f$ of $4f$ contribution for several values of the CEF splitting in the NCA and for no CEF splitting case in the NRG. The Kondo temperature defined by eq. (\ref{eq:tk}), $T_{\rm K}^{\rm (scaling)}$, and that estimated from specific heat, $T_{\rm K}^{\rm (NCA)}$, are shown.}
	\label{fig:entropy}
\end{figure}

Figure \ref{fig:entropy} shows temperature dependence of entropy $S_f$ for several values of the CEF splitting $\Delta_{\rm CEF}$.
We have taken the same values of coupling constants $J_1 \rho_c=0.2$ and $J_2 \rho_c=0$ and band width $D=10^4$K as those in ref. \citen{otsuki2}.
Without the CEF splitting, entropy tends to roughly $\ln 2$ at low temperatures. This is because only the entropy of the pseudo-spin $\mib{S}_1$ disappears due to the Kondo screening , and the pseudo-spin $\mib{S}_2$ remains free. 
On the other hand, there is no residual entropy in the cases of $\Delta_{\rm CEF}=20,\ 40$K.
This means that the fixed point of the CEF singlet is correctly reproduced in these cases. 

\begin{figure}[t]
	\begin{center}
	\includegraphics[width=0.95\linewidth]{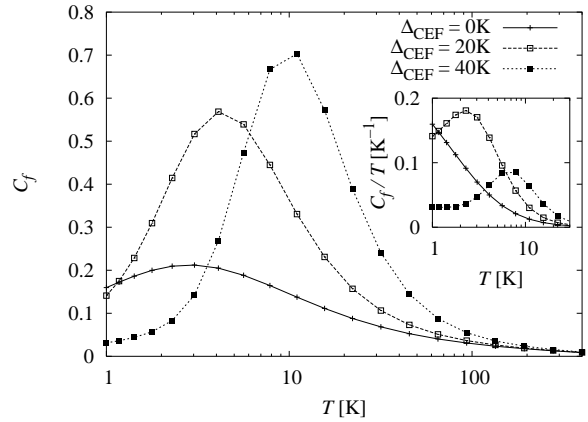}
	\end{center}
	\caption{Specific heat $C_f$ of $4f$ contribution and $C_f / T$ (inset) for several values of the CEF splitting computed in the NCA.}
	\label{fig:heat}
\end{figure}

Figure \ref{fig:heat} shows specific heat $C_f$ of $4f$ contribution in the same parameters as those in Fig. \ref{fig:entropy}.
The peak in the case of $\Delta_{\rm CEF}=0$K is due to the Kondo effect and that with $\Delta_{\rm CEF}=20,\ 40$K is due to the CEF excitations. 
We notice that the CEF splittings are about half of bare ones because of renormalization effect of the splitting.
It is consistent with the numerical results for the temperature dependence of resistivity and the dynamical magnetic susceptibility\cite{otsuki2}.

The result of specific heat $C_f$ can be utilized for an estimation of the Kondo temperature $T_{\rm K}$. 
It is known that $C_f$ become maximum at about $T_{\rm K}/3$. 
From $C_f$ without the CEF splitting (Fig. \ref{fig:heat}), we estimate as $T_{\rm K}^{\rm (NCA)} \sim 9$K, which is larger than the value estimated from the magnetic relaxation rate\cite{otsuki2}. 
On the other hand, the second order scaling theory gives 
an analytical expression of the Kondo temperature as
\begin{align}
	T_{\rm K}=D \sqrt{J \rho_c} \exp(-1/J \rho_c),
	\label{eq:tk}
\end{align}
which gives $T_{\rm K}^{\rm (scaling)} \simeq 30 {\rm K}$. 
That energy scale is consistent with the NRG result for entropy (Fig. \ref{fig:entropy}). 
Although the NCA fails to give the accurate energy scale because of the small number $n=2$ of scattering channels, 
it seems that the physical quantities are reasonably derived except for very low temperatures provided they are scaled by the alternative energy scale $T_{\rm K}^{\rm (NCA)}$. 

We now discuss application of the present theory to Pr skutterudites.
For example, PrFe$_4$P$_{12}$ shows a non-magnetic order at low temperatures\cite{PrFeP}. 
Heavy fermion states are observed in a case where the order is suppressed by applying magnetic field\cite{PrFeP} or another case diluting Pr concentration\cite{PrLaFeP}. 
Our results obtained with the impurity model are applicable to the latter case.
Applying the quartet model proposed in ref. \citen{kiss} to Pr$_x$La$_{1-x}$Fe$_4$P$_{12}$, we attribute the large value of $C_f /T$ to the Kondo effect as shown in the inset of Fig. \ref{fig:heat}. 
Namely the degeneracy of the CEF singlet-triplet levels and the Kondo screening within the quartet states are relevant to the heavy mass in Pr$_x$La$_{1-x}$Fe$_4$P$_{12}$.

In conclusion, we have presented the new set of integral equations which give the specific heat and entropy in the NCA without numerical differentiation.  It turns out the Kondo energy in the case of the singlet-triplet CEF system is underestimated in the NCA.  With proper rescaling of the energy, however, the temperature dependence of  thermodynamic quantities behaves reasonably well in the wide temperature range.

\end{document}